# Assessment of Urban Ecological Service value used in Urban Rail Transit Project

Yijie Li, Jing Chen

**Abstract:** Ecosystem services refer to the ones human beings often obtain from the natural environment ecosystem. In order to solve the problem of environmental degradation, based on **the Integrated Valuation of Ecosystem Services and Trade-offs** (**InVEST model**), this paper makes an innovation by adding the urban module that was not in the previous models, which can better deal with the evaluation of ecosystem services in urban scenarios.

## 1. Introduction

### 1.1 Background

In traditional land use projects, there is a lack of protection of the ecological environment, ignoring the impact of land use on its ecosystem services (of services obtained by human beings from natural ecosystems that directly or indirectly support human production and life) and changing in ecosystem services during utilization. In every decision of land use, due to the qualitative thinking, people lack the consideration of impact of the ecological circle, which leads to the degradation of the environment in the cumulative changes. Therefore, to reduce the economic costs associated with inappropriate land use, we have created an ecosystem service assessment model to consider ecosystems. Understanding the environmental and economic costs of land use projects in maintaining the biosphere.

## 2. Assessment of Urban Ecological Service value used in Urban Rail Transit Project

### 2.1 Introduction to selected cities

The city we choose can be expressed as City L, which is a coastal city with a length of 67.85 km from east to west, a width of 37.5km from north to south, and a length of 103.3 km from the coast. The city's economic development, employment and civic welfare depends on the products and services provided by the marine and coastal ecosystems. For instance, abundant species and genetic resources, nutrient storage and recycling, purification of land-based pollutants, stabilization of shore lines, etc. At the same time, marine ecosystem plays a key role in regulating climate and maintaining air quality, which is the main carbon and oxygen source.

## 2.2 Urban Rail Transit Project

The city's new urban rail transit project, which revolves around the entire urban area, is in line with the substance of large-scale land use projects. As City L is a coastal city, it has both terrestrial and marine ecological environment. Therefore, combining the results of marine and coastal ecosystem value assessment with the urban ecosystem value assessment, the comprehensive value of ecosystem services of the urban rail transit project is obtained. And the cost-benefit ratio analysis before and after the environmental cost is added to obtain the true comprehensive evaluation of the project.

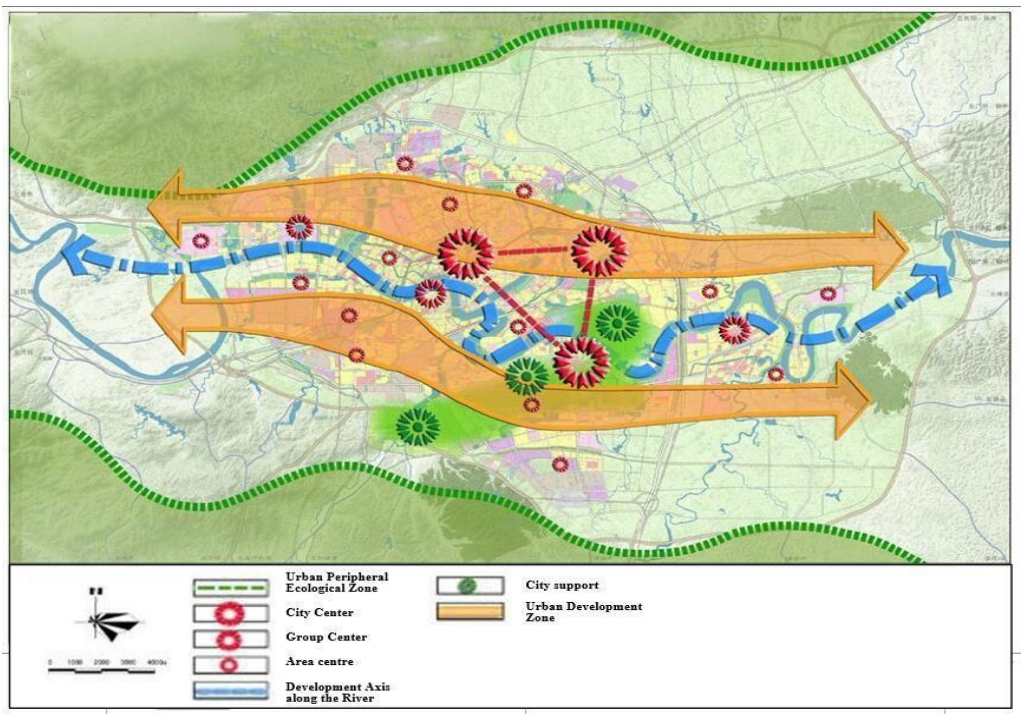

Fig 5-1 Urban rail transit hierarchical map

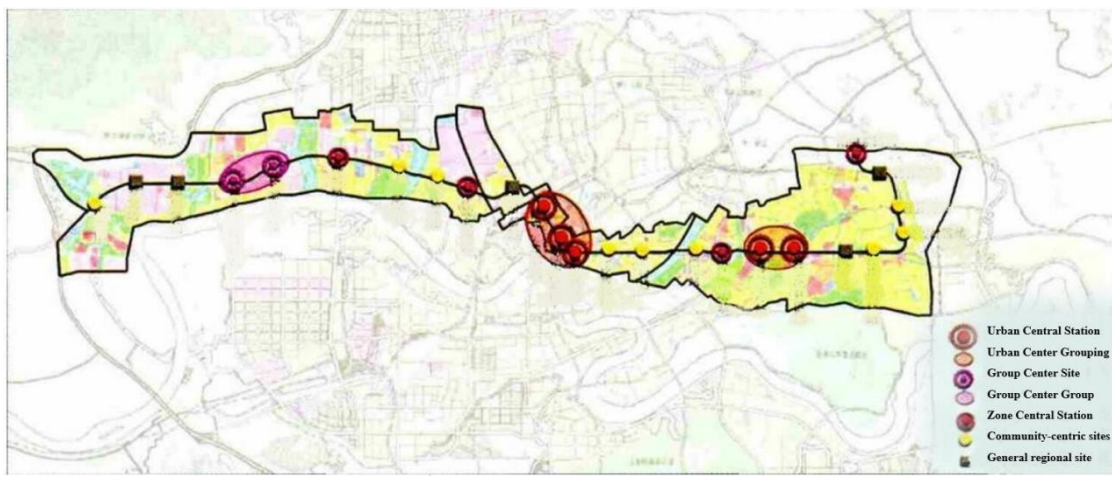

Fig 5-2 Urban rail transit process map

### 5.2.1 Assessment of marine coastal ecosystems

● Climate regulation service: the economic value of climate regulation service provided by the unit area study area of Pa. The cost of fixed $CO_2$ is recorded as $Cost_1$, and the cost of releasing $CO_2$ is $Cost_2$. The evaluation formula is as follows:

$$P_a = (1.63 Cost_1 + 1.19 Cost_2) \qquad (5\text{-}1)$$

Calculated by InVEST software, the service value of climate control in City L is 0.02($/m²•a).

● Pollution treatment and control services: the value of pollution treatment and control services per unit area is: Pev. The annual environmental capacity of the first species of pollutants in a sea area is Xi(ton/a); the treatment cost of i pollutants is

Ci($/ton); The sea area is S(m²); The value of h (m) 's annual pollution treatment and control services is Pv($/m²•a). The evaluation formula is as follows:

$$\Delta v = \sum_{i=1}^{n} Xi Ci / Q$$

$$P_v = \Delta_v (Sh)$$

$$Pev = \frac{Pv}{S} = h \sum_{i=1}^{i=n} xici / Q$$

(5-2)

Calculated by InVEST software, the service value of pollution control ecosystem is 0.60($/m²•a).

● Landscape service: set up an index Uij representing the importance of the i region; Denote the j activity or the use of the j landscape in the i region (1 for use and 0 for non-utilization). The evaluation formula is as follows:

$$IS_i = \sum_{ij} U_{ij} I_j \qquad (5\text{-}3)$$

The value of landscape ecosystem services calculated by InVEST software is 0.11($/m²•a).

● Service of fishery resources: the value of serving fishery resources Pmf($/m²•a); The annual cost of marine fishing Cmf($/m²•a); The area of marine fishing S(m²). The evaluation formula is as follows:

$$P_{mf} = \frac{\overline{R_{mf} - C_{mf}}}{S} \qquad (5\text{-}4)$$

Calculated by InVEST software, the average value of ecosystem services for fishery resources is 0.32($/m²•a).

### 5.2.2 Assessment of urban ecosystems

Through the weights determined by the entropy weight method introduced in the previous section, we derive the calculation formula for the assessment of urban ecosystem services:

$$P_{urban} = \frac{P_v}{S} \bullet \frac{P_0 E}{\delta} \sigma P_s \rho \qquad (5\text{-}5)$$

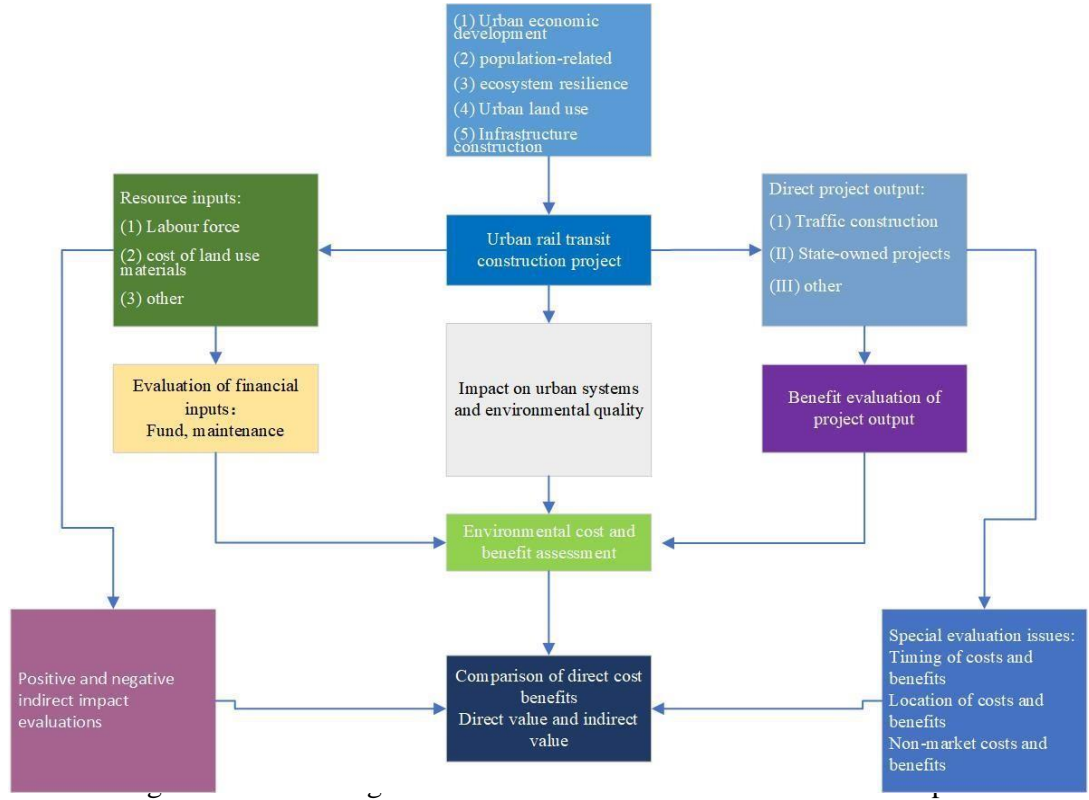

Fig 5-3 Urban ecological service assessment model establishment process

● $\rho$ is the level of ecological value assessed by entropy weight method and fuzzy evaluation model, which is equal to the specific monetary value of other ecological value evaluation in InVEST model.

(1) Using Fuzzy Comprehensive Evaluation to determine Evaluation Formula:

$$\theta = W \bullet R_{M \times N} \qquad (5\text{-}6)$$

$\theta$ is the result of the model evaluation, and W = ($w_1, w_2, w_3, w_4, w_5$) is the five general urban evaluation indicators (urban economic development, population correlation and distribution, ecosystem resilience, etc.) mentioned in the previous section. The comprehensive weight of urban land use and infrastructure construction, R is the relation matrix of subjection degree of each evaluation index to the standard matching.

$$R = \begin{pmatrix} R_{11} & R_{12} & R_{13} & R_{14} & R_{15} \\ R_{21} & R_{22} & R_{23} & R_{24} & R_{25} \\ R_{31} & R_{32} & R_{33} & R_{34} & R_{35} \\ R_{41} & R_{42} & R_{43} & R_{44} & R_{45} \\ R_{51} & R_{52} & R_{53} & R_{54} & R_{55} \end{pmatrix} \qquad (5\text{-}7)$$

(2) Concrete example calculation

By introducing the latest data from City L in 2019 into the evaluation model, the W and R matrix is obtained as follows:

$$R = \begin{pmatrix} 0.7723 & 0.5383 & 0.5443 & 0.032 & 0.733 \\ 0.0024 & 0 & 0.7742 & 0.042 & 0.134 \\ 0.8932 & 0.2234 & 0.2574 & 0.045 & 0.356 \\ 0.3334 & 0.1595 & 0.1241 & 0.024 & 0.251 \\ 0.1672 & 0.4325 & 0.0004 & 0.234 & 0.001 \end{pmatrix} \quad (5\text{-}8)$$

$$W = (0.452 \quad 0.675 \quad 0.986 \quad 0.463 \quad 0.523) \quad (5\text{-}9)$$

$$\theta = W \bullet R = 0.5637 \quad (5\text{-}10)$$

$$P_{urban} = \frac{P_v \bullet P_0 E}{S} \bullet \sigma P \rho$$

$$= \frac{3.897}{0.324} \bullet \frac{0.235 \bullet 10^s}{3.213} \bullet 0.0721 \bullet 2.232 \bullet 0.542 \quad (5\text{-}11)$$

$$= 0.56(\$/m^2 \bullet a)$$

E is the original related cost of environmental protection in urban planning, $\sigma$ is viscosity coefficient ($0 < \sigma < 1$). $P_0$ is the comprehensive productivity of urban land per unit area.

### 5.2.3 Analysis of Project Cost-benefit by adding Environmental cost

#### 5.2.3.1 Cost analysis

● **Determining cost unit**

Our team divided the cost into two categories: tangible cost and intangible cost.

● **Tangible cost** —— Tangible costs involved in the construction of urban rail transit

Ⅰ. Expenditures on investment-related commodities and raw materials.

Ⅱ. Operating expenses of urban rail transit.

Ⅲ. Real estate cost.

Ⅳ. Insurance, wages and taxes paid of employee.

Ⅴ. Maintenance of electricity and water charges.

**Intangible cost**

Ⅰ. Time spent on projects (benefits from other projects spent on dunes).

Ⅱ. External energy for project operation.

Ⅲ. Assessment of possible business losses during project operation.

Ⅳ.Project environmental costs resulting from the value assessment of ecosystem services.

**5.2.3.2 Benefit analysis**

**Direct value:** it facilitates people's transportation and improves people's demand efficiency; urban rail transit construction can also increase tax revenue and contribute to the country.

**Indirect value:** taking the function provided by the ecosystem in the City as an example

**The value of choice:** the construction along the urban rail transit can be regarded as a new type of value, such as the landscape and tourist areas established along the line will benefit more because of the urban rail transit.

**5.2.3.3 Conclusion**

The cost benefit ratio of urban rail transit project before and after the addition of environmental cost was calculated by analyzing the cost and benefit of the urban rail transit project.

It is calculated that the cost-benefit ratio of the project before adding environmental cost is 0.583, and that after adding environmental cost is 0.696. Although the original cost increases after adding environmental cost, the overall benefit is increased. This reflects the need to add environmental costs to the total cost of the project.

# 3. The implications of modeling on planners and managers

## 3.1 Planners and managers

The schematic diagram of the project planning management system is shown in the Fig 6-2. General project management system consists of project process modeling tools, project management, task table manager, user interface and related applications and data.

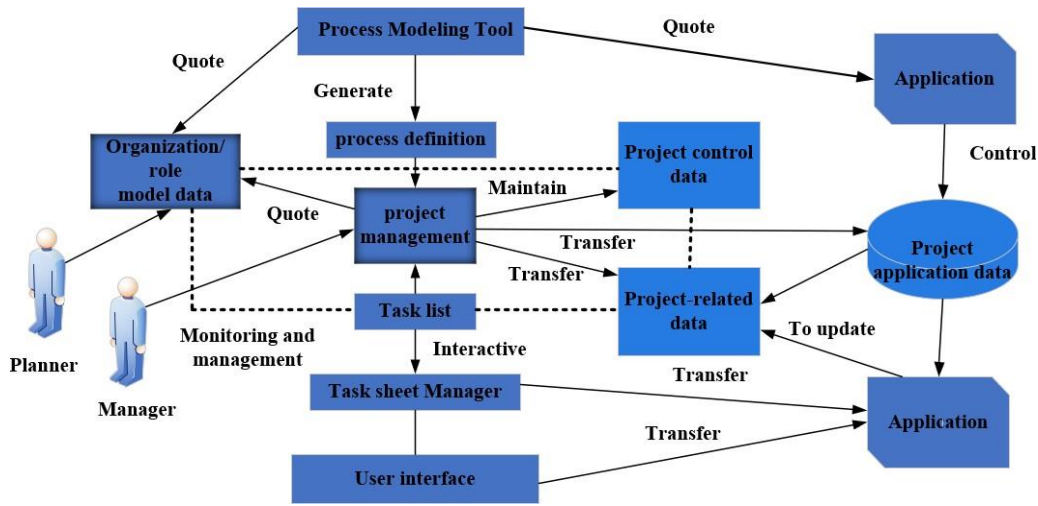

Fig 6-1 Structural Chart of Project Planning Management System

As shown in the Fig 6-2, the implementation of management system is generally divided into three stages.

Modeling stage. Based on the analysis of a specific application, the enterprise business process model is established by using the modeling tool, and the actual business process of the enterprise is transformed into a computer-processable model.

Model instantiation phase. According to the application project, the parameters required for each process are set and the resources required for each activity are allocated.

Model execution phase. To complete the execution of the business process, it is necessary to abide by the rules defined in the model establishment stage.

Then, it includes process rules, activity rules and organization definition rules to complete the implementation of human-computer interaction and application.

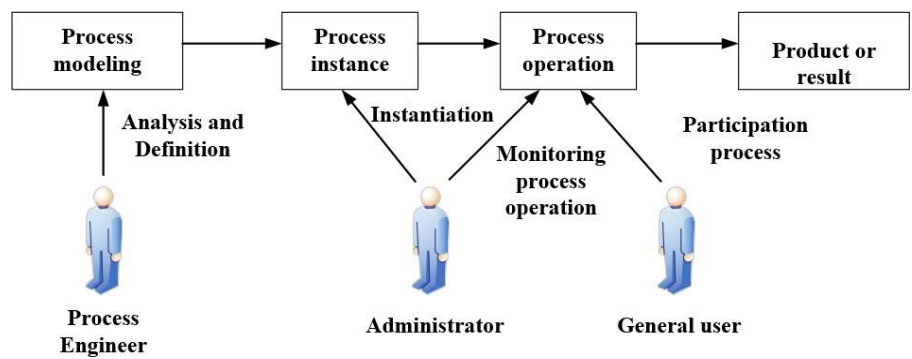

Fig 6-2 Three phases of project management implementation

## 3.2  Implications

From the analysis of charts and data, we can see that the value of ecosystem services in L city is decreasing as a whole. In addition, due to the diversity of coastal zone organisms, the service value of climate regulation and maintenance of air quality and the service value of flood control, the Moisture-proof and the stability of coastline are higher.

From the analysis results, we can clearly realize that although adding environmental costs may increase the original budget, the sustained benefits of ecological services are considerable. If the planning is unreasonable and the ecosystem is not protected and maintained, it may be difficult to recover the damage to the local ecosystem caused by the project itself. The most direct consequence is to reduce the value of ecosystem services, which is the cost of degradation.

Through the above analysis, it can be concluded that when the value of urban ecosystem services is obtained based on this model, planners must pay attention to the value of ecosystem, fully investigate and evaluate the value of ecological services around the project construction, add environmental costs, and adjust part of the details of specific operations of the project in real time according to the change of the environment. For managers, the necessary maintenance of the ecological environment should be monitored and regularly carried out in order to maintain the current benefits of environmental costs.

## 4. Long-term adaptive dynamic assessment model

### 4.1 Adjustment of model over time

From our model, we can get several main evaluation factors, such as urban economic development, population correlation and distribution, ecosystem resilience, urban land use, infrastructure construction, to construct the evaluation system of urban ecological services.

As for the development of urban economy, the following factors will affect it: land, population, industry, science and technology, culture and policy. Among these six elements, the first three elements are visible, while the last three elements are invisible. In other words, they are abstract and can only be shown through other objects.

Among the six factors, industry is the most direct factor to promote urban economic development, while the other five factors are indirect, and ultimately affect industry to promote economic development. Man is the most important factor among the six factors. The importance lies in that man is the main participant and the enjoyer of the results of social development. Without human participation, the economic development of cities is meaningless, and it is impossible to have a so-called city at all. The combination of six elements plays an important role in the development of urban economy. The direct driving force of urban economic development lies in industry. The other five factors all play an indirect role and influence each other. (Picture) For the first three elements, we calculate the new weights of each factor based on the new data obtained in the next year and modify the model parameters to adapt to the new changes based on the new weights. For the latter three factors, due to indirect effect, it is impossible to find out the specific value, and when the value occurs, we may know its specific impact. But each one can have a significant impact. Therefore, when a place's culture may not change in a short time, but the promotion of science and technology, policy changes and other factors we need to consider, the model also needs to change.

Implementing ecological construction and protection projects, actively improving the ecological environment, enhancing the overall awareness of environmental protection, promulgating relevant environmental protection policies, natural disasters, large influx of immigrants, industrial upgrading and so on, all need to adjust and change the model at this time. It can be seen that the model must be highly flexible. In short, only dynamic models can meet the needs of the actual situation.

## 4.2 Time Prediction Model Based on LSTM

Long-short term memory [1] is a special RNN [2,3] model, which is proposed to solve the problem of gradient dispersion in RNN model. Among them, the internal state of RNN network can show the dynamic sequential behavior. Unlike feedforward neural networks, RNN can use its internal memory to process input sequences of arbitrary time series, which makes it easier to process such as non-segmented handwriting recognition, speech recognition and so on. However, RNN has two problems, namely, gradient disappearance and gradient explosion. The original intention of LSTM design is to solve the problem of long-term dependence in RNN, so that remembering long-term information becomes the default behavior of neural network, rather than a lot of effort to learn. LSTM model replaces RNN cells in the hidden layer with LSTM cells to make them have long-term memory ability. At present, the popular LSTM model structure is Cho, et al [4], which uses the forgetting gate and the input gate to receive and input parameters respectively.

In the LSTM neural network model, the forward calculation method can be expressed as:

$$f_t = \sigma(W_f[h_{t-1}, X_t] + b_f) \tag{7-1}$$

$$i_t = \sigma(W_i[h_{t-1}, X_t] + b_i) \tag{7-2}$$

$$\tilde{C}_t = tanh(W_c[h_{t-1}, X_t] + b_c) \tag{7-3}$$

$$C_t = f_t * C_{t-1} + i_t * \tilde{C}_t \tag{7-4}$$

$$o_t = \sigma(W_o[h_{t-1}, X_t] + b_o) \tag{7-5}$$

$$h_t = o_t * \tanh(C_t) \tag{7-6}$$

Among them, f, i, c, o respectively represent forgetting gate, output gate, cell state, output gate, W and b are corresponding weight coefficient matrix and bias term. Rou and tanh are sigmoid and hyperbolic tangent activation functions respectively. From the introduction of the formula, we can see that LSTM can solve the long-term dependence problem in RNN. From the introduction of the formula, we can see that LSTM can solve the long-term dependence problem in RNN. Therefore, we can build a neural network prediction model based on LSTM to predict the value of different ecosystem services in the coming years.

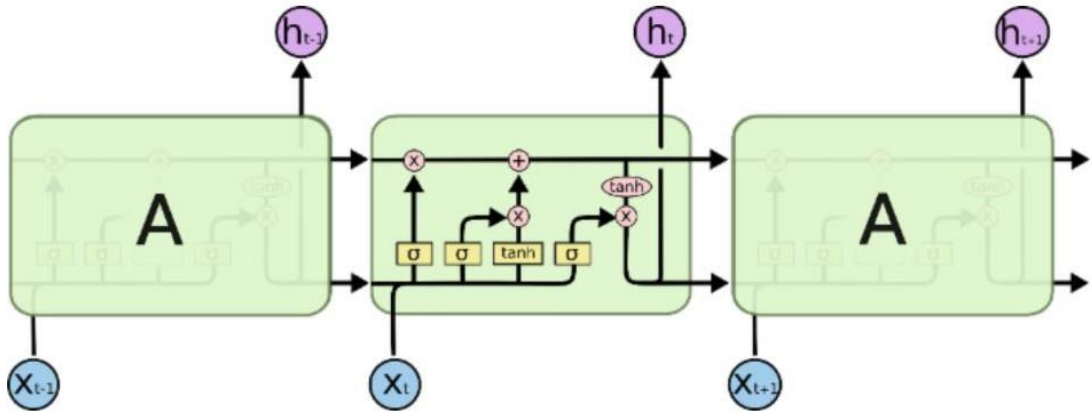

Fig 7-1 The repetitive module in LSTM consists of four layers of interactive neural network layer

The predicted results are shown in the Fig7-2. We can conclude that the value of ecosystem services in this area is decreasing as a whole.

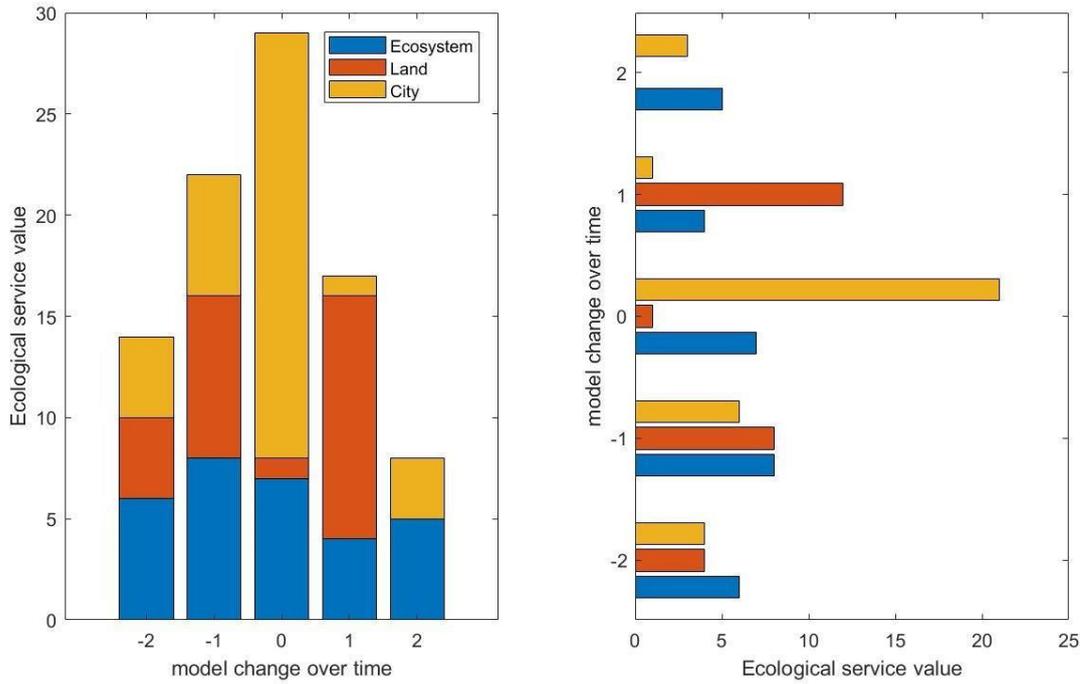

Fig 7-2 Time Prediction Model Generation Graph

## 5. Sensitivity analysis

In order to determine the effect of the impact of land use projects on the measurement of the measured value of the cost-benefit assessment index for maintaining the ecosystem, in order to determine the impact of the error on the evaluation results, Our team uses sensitivity analysis to discuss the stability of the evaluation results, and corrects the model based on the results. There are two kinds of sensitivity analysis, one is local sensitivity analysis, the other is global sensitivity analysis. Our team uses the method of global sensitivity analysis to analyze the stability of evaluation results. The value of all evaluation indicators is increased and decreased by 10 percent, and the value of the evaluation index is calculated （According to formula 7-1and 7-2） after the change of the value of the evaluation index. Maximum comprehensive correlation degree.

$$\Psi^m = \{p \mid 0 <= p_i <= 1; i = 1,2,...,m\} \tag{8-1}$$

$$g(p_1, p_2, ..., p_m) = g_0 + \sum_{i=1}^{m} g_i(p_i) + \sum_{1<=i<j<=m} g_{i,j}(p_i, p_j) + ... + g_{1,2,...k}(p_1, p_2, ..., p_m) \tag{8-2}$$

The results obtained from sensitivity analysis are consistent with those of the model, which indicates that the impact of the model on land use projects is accurate in terms of maintaining the cost-effectiveness of ecosystems. It has good robustness and adaptability.